\documentclass[
pre,
twocolumn,
superscriptaddress,
floatfix,
longbibliography
]{revtex4-1}

\usepackage[realmainfile]{currfile}
\usepackage{xr-hyper}
\makeatletter

\newcommand*{\addFileDependency}[1]{
\typeout{(#1)}
\@addtofilelist{#1}
\IfFileExists{#1}{}{\typeout{No file #1.}}
}\makeatother

\newcommand*{\myexternaldocument}[1]{%
\externaldocument{#1}%
\addFileDependency{#1.tex}%
\addFileDependency{#1.aux}%
}

\myexternaldocument{supp}

\usepackage{color}

\usepackage{graphicx}
\usepackage{epsfig,verbatim,enumerate}
\usepackage{amssymb,amsmath}
\usepackage{ifthen}

\usepackage{longtable}

\usepackage{mathtools}

\newboolean{twocolswitch}

\newcommand{\sindex}[1]{}
\newcommand{\nindex}[1]{}

\newcommand{\www}[1]{\url{#1}}

\usepackage{lettrine}

\usepackage[dvipsnames]{xcolor}

\usepackage{xifthen}
\newcommand{\dummycite}[2][]{\ifthenelse{\isempty{#2}}{\textcolor{NavyBlue}{[CITE]}}{\textcolor{NavyBlue}{[CITE\textemdash #2]}}}

\usepackage{xifthen}
\newcommand{\dummyfigure}[2][]{\ifthenelse{\isempty{#2}}{\textcolor{NavyBlue}{[FIGURE]}}{\textcolor{NavyBlue}{[FIGURE\textemdash #2]}}}

\usepackage{xifthen}
\newcommand{\dummytable}[2][]{\ifthenelse{\isempty{#2}}{\textcolor{NavyBlue}{[TABLE]}}{\textcolor{NavyBlue}{[TABLE\textemdash #2]}}}

\usepackage{soul}
\setstcolor{red}

\setboolean{twocolswitch}{true}

\renewcommand{\selectlanguage}[1]{}

\begin{document}

\title{\protect
Collective sleep and activity patterns of college students from wearable devices
}

\author{
\firstname{Mikaela Irene}
\surname{Fudolig}
}
\email{mikaela.fudolig@uvm.edu}

\affiliation{
    Vermont Complex Systems Center,
    MassMutual Center of Excellence for Complex Systems and Data Science,
    University of Vermont,
    Burlington, Vermont, USA 05405
}

\affiliation{
    Department of Mathematics and Statistics,
    University of Vermont,
    Burlington, Vermont, USA 05405
}

\affiliation{
    Computational Story Lab,
    MassMutual Center of Excellence for Complex Systems and Data Science,
    University of Vermont,
    Burlington, Vermont, USA 05405
}

\author{
\firstname{Laura S. P.}
\surname{Bloomfield}
}

\affiliation{
    Gund Institute for Environment,
    University of Vermont,
    Burlington, Vermont, USA 05405}

\affiliation{
    Vermont Complex Systems Center,
    MassMutual Center of Excellence for Complex Systems and Data Science,
    University of Vermont,
    Burlington, Vermont, USA 05405
}

\affiliation{
    Department of Mathematics and Statistics,
    University of Vermont,
    Burlington, Vermont, USA 05405
}

\affiliation{
    Computational Story Lab,
    MassMutual Center of Excellence for Complex Systems and Data Science,
    University of Vermont,
    Burlington, Vermont, USA 05405
}

\author{
\firstname{Matthew}
\surname{Price}
}

\affiliation{
  Department of Psychological Science,
  University of Vermont,
	Burlington, Vermont, USA 05405}

\affiliation{
    Vermont Complex Systems Center,
    MassMutual Center of Excellence for Complex Systems and Data Science,
    University of Vermont,
    Burlington, Vermont, USA 05405
}

\author{
\firstname{Yoshi M.}
\surname{Bird}
}
\affiliation{
  Vermont Complex Systems Center,
  MassMutual Center of Excellence for Complex Systems and Data Science,
  University of Vermont,
  Burlington, Vermont, USA 05405
  }

\affiliation{
    Computational Story Lab,
    MassMutual Center of Excellence for Complex Systems and Data Science,
    University of Vermont,
    Burlington, Vermont, USA 05405
}

\author{
\firstname{Johanna E.}
\surname{Hidalgo}
}

\affiliation{
  Department of Psychological Science,
  University of Vermont,
	Burlington, Vermont, USA 05405}

\author{
\firstname{Julia}
\surname{Kim}
}

\affiliation{
  Project LEMURS,
  University of Vermont,
	Burlington, Vermont, USA 05405}

\affiliation{
  Department of Electrical and Biomedical Engineering,
  University of Vermont,
	Burlington, Vermont, USA 05405}

\author{
\firstname{Jordan}
\surname{Llorin}
}

\affiliation{
  Project LEMURS,
  University of Vermont,
	Burlington, Vermont, USA 05405}

\affiliation{
  Department of Electrical and Biomedical Engineering,
  University of Vermont,
	Burlington, Vermont, USA 05405}

\author{
\firstname{Juniper}
\surname{Lovato}
}

\affiliation{
    Vermont Complex Systems Center,
    MassMutual Center of Excellence for Complex Systems and Data Science,
    University of Vermont,
    Burlington, Vermont, USA 05405
}

\affiliation{
    Computational Story Lab,
    MassMutual Center of Excellence for Complex Systems and Data Science,
    University of Vermont,
    Burlington, Vermont, USA 05405
}

\author{
\firstname{Ellen W.}
\surname{McGinnis}
}

\affiliation{
Wake Forest University School of Medicine,
Winston-Salem, North Carolina, USA 27101
}

\author{
\firstname{Ryan S.}
\surname{McGinnis}
}

\affiliation{
Wake Forest University School of Medicine,
Winston-Salem, North Carolina, USA 27101
}

\author{
\firstname{Taylor}
\surname{Ricketts}
}

\affiliation{
    Gund Institute for Environment,
    University of Vermont,
	Burlington, Vermont, USA 05405}

\affiliation{
    Rubenstein School of Environment and Natural Resources,
    University of Vermont,
	Burlington, Vermont, USA 05405}

\author{
\firstname{Kathryn}
\surname{Stanton}
}

\affiliation{
  Project LEMURS,
  University of Vermont,
	Burlington, Vermont, USA 05405}

\affiliation{
  Department of Electrical and Biomedical Engineering,
  University of Vermont,
	Burlington, Vermont, USA 05405}

\author{
\firstname{Peter Sheridan}
\surname{Dodds}
}

\affiliation{
    Vermont Complex Systems Center,
    MassMutual Center of Excellence for Complex Systems and Data Science,
    University of Vermont,
    Burlington, Vermont, USA 05405
}

\affiliation{
    Department of Computer Science,
    University of Vermont,
    Burlington, Vermont, USA 05405
 }

\affiliation{
    Computational Story Lab,
    MassMutual Center of Excellence for Complex Systems and Data Science,
    University of Vermont,
    Burlington, Vermont, USA 05405
}

\affiliation{
  Santa Fe Institute,
  New Mexico, USA 87501
}

\author{
\firstname{Christopher M.}
\surname{Danforth}
}

\affiliation{
    Vermont Complex Systems Center,
    MassMutual Center of Excellence for Complex Systems and Data Science,
    University of Vermont,
    Burlington, Vermont, USA 05405
}

\affiliation{
    Department of Mathematics and Statistics,
    University of Vermont,
    Burlington, Vermont, USA 05405
}

\affiliation{
    Computational Story Lab,
    MassMutual Center of Excellence for Complex Systems and Data Science,
    University of Vermont,
    Burlington, Vermont, USA 05405
}


\begin{abstract}
  \protect
  To optimize interventions for improving wellness, it is essential to understand habits, which wearable devices can measure with greater precision.
  Using high temporal resolution biometric data taken from the Oura Gen3 ring, we examine daily and weekly sleep and activity patterns of a cohort of young adults ($N=582$) in their first semester of college. 
A high compliance rate is observed for both daily and nightly wear, with slight dips in wear compliance observed shortly after waking up and also in the evening.
Most students have a late-night chronotype with a median midpoint of sleep at 5AM, with males and those with mental health impairment having more delayed sleep periods.
Social jetlag, or the difference in sleep times between free days and school days, is prevalent in our sample.
While sleep periods generally shift earlier on weekdays and later on weekends, sleep duration on both weekdays and weekends is shorter than during prolonged school breaks, suggesting chronic sleep debt when school is in session.
Synchronized spikes in activity consistent with class schedules are also observed, suggesting that walking in between classes is a widespread behavior in our sample that substantially contributes to physical activity.
Lower active calorie expenditure is associated with weekends and a delayed but longer sleep period the night before, suggesting that for our cohort, active calorie expenditure is affected less by deviations from natural circadian rhythms and more by the timing associated with activities.
Our study shows that regular sleep and activity routines may be inferred from consumer wearable devices if high temporal resolution and long data collection periods are available. 

\end{abstract}

\maketitle

\section{Introduction} 

Deleterious effects on wellness can be due not just to individual choices but also to social contexts such as work~\cite{jansen_work_2003, bushnell_work_2010, flo_shift-related_2013, akerstedt_work_2015, ma_relationship_2018} and academic schedules~\cite{malheiros_school_2021, wu_association_2023}.
To develop interventions that will work with or seek to revise such contexts, we need to understand how they relate to the habits they influence.
These behaviors have been studied for many populations~\cite{carney_daily_2006, olds_sleep_2011, moss_is_2015, holding_sleepiness_2020, sinha_impact_2020, gupta_changes_2020, suckow_daily_2023}, including college students ~\cite{galambos_losing_2009, lund_sleep_2010, orzech_state_2011, onyper_class_2012}, using surveys which are susceptible to recall error~\cite{coughlin_recall_1990} and social desirability bias~\cite{van_de_mortel_faking_2020}.

The use of digital data to examine these behaviors addresses limitations associated with self-reports. Data on mobile phone activity~\cite{eagle_eigenbehaviors_2009, monsivais_tracking_2017, schoedel_challenge_2020}, mobile device location~\cite{farrahi_what_2008, sevtsuk_does_2010, jo_spatiotemporal_2012}, social media usage~\cite{linnell_sleep_2021, zhou_how_2023} and wearable sensors ~\cite{purta_experiences_2016, sano_measuring_2016, lin_social_2019, wu_multi-modal_2021} can reflect daily routines associated with biological factors, such as circadian rhythms, and social factors, such as work hours. The growing availability of wearable devices opens an avenue to relate these routines to physiological measures. Perhaps due to the potential of personalized healthcare that wearable devices may offer, most studies on wearable devices focus on the individual. In contrast, there is much less research on how wearable devices can indicate group behavior resulting from a shared context~\cite{wirz_decentralized_2009, gordon_recognizing_2012, li_multi-user_2020}. Fine temporal resolution sleep and activity data measured by wearable devices have been underutilized in studying behavior~\cite{karimi_longitudinal_2024}, offering opportunities for novel, nuanced observations using these methods.

Adolescents who are transitioning to college are a vulnerable population characterized by a developing executive functioning system~\cite{chung_transitional_2017} paired with new and plentiful opportunities to make decisions without caregiver input. This period of transition is marked by high rates of mental health impairment associated with substance misuse, morbidity, and mortality~\cite{jones_adult_2013, pedrelli_college_2015, bruffaerts_mental_2018, oswalt_trends_2020}.
As students adapt to the novel social contexts of the college environment, they can form new habits~\cite{marien_studying_2019} that can improve or degrade their well-being~\cite{copeland_daily_2022}.
To understand the wellness behaviors and habits of and ultimately develop engaging interventions for this age group, the Lived Experiences Measured Using Rings Study (LEMURS)~\cite{price_large_2023, bloomfield_predicting_2024, fudolig_two_2024} observed behaviors among first-year college students using data from both surveys and the Oura Gen3 ring, a wearable device that monitors sleep and physical activity. All participants were recruited from the same class year at a single university over an 8-week study period creating a controlled, comparable shared context that shaped daily routines. As a result, our dataset is particularly well suited to determine how collective behavior manifests via wearable devices.

We address gaps in the literature by using fine temporal resolution data from the Oura ring to reconstruct daily sleep and activity patterns for both free days and school days (weekdays vs. weekends, school in session vs. school breaks).
We also look at if and how school disrupts natural sleep habits, which reflect an individual's chronotype~\cite{roenneberg_life_2003}, that are observed when school is not in session.
This disruption is termed \textit{social jetlag}~\cite{wittmann_social_2006, roenneberg_chronotype_2019}, so-called because it is similar to the change in sleep patterns after moving time zones, except that it is due to social conventions rather than travel.
The activity data from the Oura ring is used to infer collective routines and also provides the active calorie expenditure.
We then compare how these routines differ across groups of students.
Specifically, we look at gender and reported impairment due to anxiety or depression, with these two conditions being of interest due to their prevalence in our sample.
All findings are then interpreted in the context of the academic schedule.

\section{Methods}
\label{sec:methods}

Under LEMURS, first-year college students recruited from a university in the northeastern United States were asked to wear the Oura Gen3 ring at all times except when charging or doing activities during which it was uncomfortable to wear (e.g., lifting heavy weights).
They also completed surveys, including a baseline survey on their demographic information and mental health indicators.
LEMURS was conducted over 8 weeks (Oct--Dec 2022) within a single semester and included the Thanksgiving break, a full week during which school was not in session.
There were $N=582$ individuals who completed the baseline survey and had both sleep and activity data from the Oura ring for at least one day.
Of these individuals with sleep and activity data, 65\% were female, 28\% were male, and 6\% were non-binary individuals. These percentages are representative of the entire first-year undergraduate student population at the university.
35\% reported impairment due to diagnosed anxiety or depression.

We consider the following data provided by the Oura ring: the start and end times for the longest nightly sleep period starting between 6 PM and 6 AM; active calories expended per day; and the time series of physical activity level categories recorded at 5-minute intervals from 4 AM to 4 AM the next day.
These categories are based on the metabolic equivalents of task (METs), the ratio of energy expenditure for a given activity to that when resting. 
The Oura ring reports the following activity level categories~\cite{lomas_defining_2023, oura_support_how_nodate} based on METs as integers: 1=rest ($\textrm{MET}<1.05$), 2=inactive ($1.05 < \textrm{MET} < 2$), 3=low, 4=medium, and 5=high.
Typical medium-intensity activities are jogging, hiking, and rollerskating, during which the heart rate increases, but one can still continue a conversation.
High-intensity activities like running, skiing, and cycling cause higher increases in heart rate and respiratory demand~\cite{oura_support_daily_nodate}.
In addition to these categories, periods of non-wear are also reflected in the activity level time series.

\subsection{Ring wear compliance}

We first examine participant compliance in wearing the ring.
At a daily level, we look for the number of users with sleep or activity data for a given date.
As the 5-minute activity time series also records periods of non-wear, we can compute the number of days a user has worn the ring at a given time.

\subsection{Weekly patterns in physical activity and sleep}

We examine sleep patterns, focusing on the timestamps of the start and end of the sleep period measured by the Oura ring.
By convention, we refer to the start and end of the sleep period in the rest of the paper as ``bedtime'' and ``get-up time'', respectively.
For each user, we consider days of the week (Monday--Sunday) with three (3) or more weeks of activity data, excluding Thanksgiving week.
This reduces the number of users examined to $N=572$. However, the proportions by gender and mental health impairment remain unchanged.
All days of the week also have a similar number of individuals that meet this criterion (Table S1).

We then obtain the activity level per user for each 5-minute interval, which we average across all weeks for every day of the week, disregarding non-wear periods in taking the mean.
The mean of the resulting per-user mean activity time series is then taken as the collective activity time series.
We compare collective activity with the per-user mean bedtimes and get-up times observed for every day of the week.

We also compare the sleep and activity patterns by gender and impairment due to anxiety or depression using statistical parametric mapping (SPM)~\cite{friston_statistical_1994}.
SPM, originally developed for functional neuroimaging, first assembles the result of individual statistical tests for every voxel.
It then uses random field theory to set statistical significance that controls for Type 1 errors taking into account spatial correlations.
In analyzing the collective activity time series, each time point is treated like a voxel in 1D space.
Using SPM in conjunction with pairwise t-tests, we obtain the time intervals in a day where we observe differences in the activity patterns statistically significant at $\alpha=0.05$. 
SPM calculations are implemented with the Python module \texttt{spm1d}~\cite{pataky_one-dimensional_2012}.

\subsection{Social jetlag and activity}

A quantitative indicator of chronotype is the midpoint between sleep onset and wake-up time, a measure derived from the Munich Chronotype Questionnaire~\cite{roenneberg_life_2003}.
Social jetlag~\cite{roenneberg_chronotype_2019} is the difference in the midpoint of sleep on free days (MSF) and school or work days (MSW): $SJL=MSF-MSW$.
Note that social jetlag can be positive or negative, and that both indicate a deviation from the natural sleep habits observed during free days.
While weekends have been traditionally used to compute the MSF, sleep debt accumulated during the weekdays tends to be recovered during the weekends, making the weekend measure inaccurate.
While corrections using more complicated formulas have been proposed~\cite{roenneberg_chronotype_2019}, we instead compute the MSF as the average of the sleep period midpoints from Thanksgiving week.
We obtained the chronotypes for $N=541$ individuals, with the proportions by gender (66\% female, 28\% male and 6\% non-binary) and impairment status (34\% with impairment, 66\% without) almost unchanged.
The social jetlag for a sleep period outside of Thanksgiving week is then defined as the difference between the MSF and the midpoint of that sleep period.

We examine the association between the MSF with gender and mental health impairment.
The MSF is recorded as the number of hours beyond 12MN; if the midpoint is earlier than 12MN, this number is negative.
The MSF is used in a linear regression model as the response variable and the gender and mental health impairment status (present or absent) of the individual as the predictors.
Note that changes in the midpoint of sleep can be due to differences in any combination of the start time and end time of the sleep period as well as the sleep period duration.
To understand which of these is affecting the MSF, we take the per-user means of the sleep period start times, the sleep period end times, and the sleep period durations during Thanksgiving week and use each of these as lone predictors in a linear regression model for predicting the MSF.

Unlike the MSF, for which there is only one value for each participant, social jetlag varies across sleep periods for an individual.
We therefore use linear mixed-effects models to understand how social jetlag is associated with gender, mental health impairment, and the day of the week (weekday vs. weekend) with the participant ID and the week number as random effects.

Similarly, we use linear mixed-effects models with the log-transformed active calorie expenditure as the response and social jetlag, bedtime duration, gender, mental health impairment and day of the week (weekday vs. weekend) as predictors.
Zero calorie values were replaced with a value of 0.1 for the log transform calculation (Figure S8).
Again, the participant ID and the week number are random effects.
For these models, we exclude Thanksgiving week, the reference point for social jetlag.

All these models used $\alpha=0.05$ as the threshold for statistical significance.
Regressions were implemented using the Python package \texttt{pymer4}~\cite{jolly_pymer4_2018}.

\section{Results}
\label{sec:results}

\subsection{Ring wear compliance}

\begin{figure}[ht!]
    \centering
    \includegraphics[width=\columnwidth]{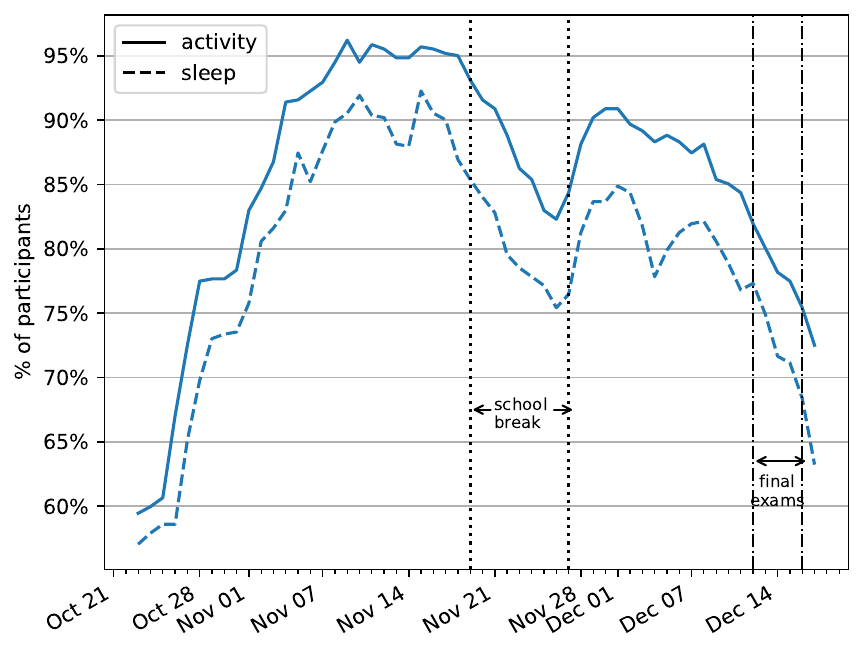}
    \caption{\textbf{Ring wear compliance of users over the study period.} We observe at least 70\% of the participants with sleep and physical activity data for most of the study period. Declines in participation are observed during the Thanksgiving break and the final exam period indicated in the figure by vertical lines.}
    \label{fig:compliance-studyperiod}
\end{figure}

We observe high rates of compliance for both sleep and activity data, with at least 70\% of the participants wearing the ring during both daytime and nighttime except for the first and last few days (Figure~\ref{fig:compliance-studyperiod}).
There is a sharp increase in ring wear within the first two weeks as participants begin to wear or get used to wearing the ring, while declines are observed during Thanksgiving break and the final exam period.
We note that depending on their schedule, students may end their semester before the last day of the final exam period.
Outside of sleep, participants also consistently wear their rings most of the time, with the median percentage of days in the study period with at least 12 hours of non-rest wear time being 77\%.
High compliance is observed for both weekdays and weekends.
Similar patterns are observed for different genders and for those with or without impairment due to anxiety or depression (Figures S1 and S2).

\begin{figure}[ht!]
    \centering
    \includegraphics[width=\columnwidth]{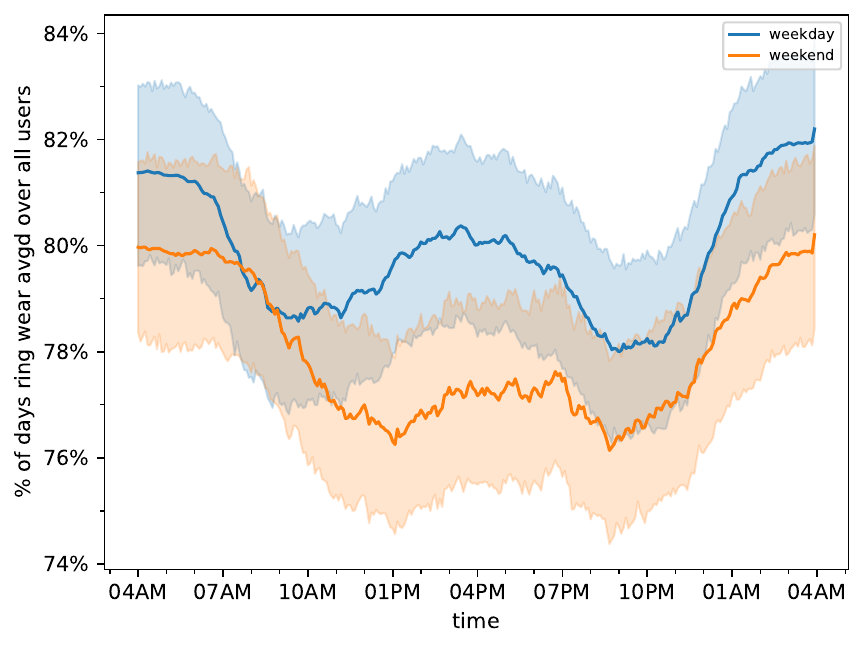}
    \caption{\textbf{Ring wear compliance over the course of a day.} For each user, we take the percentage of weekdays and weekends (excluding the week of Thanksgiving break) that they wear the ring at a given time. The solid curves show the average wear compliance of the time series across all users, and the shaded areas indicate the 95\% confidence interval. }
    \label{fig:compliance-hourly}
\end{figure}

Using the activity time series taken every 5 minutes, we can construct the wear pattern of users within a day, particularly how often a user wears the ring at a given time.
Non-wear may be attributed to charging the ring or performing activities that users prefer to do without wearing the ring.
While wear compliance is generally high over the course of a day, dips in wear consistency are observed in the morning and evening. 
In contrast to the consistent non-wear period in the evening, the non-wear period in the morning is markedly different for school days and non-school days (Figures~\ref{fig:compliance-hourly} and S3).
While the evening non-wear period is similar for weekdays and weekends, the morning non-wear period is shifted later in the day during weekends as compared to weekdays.

\subsection{Patterns in sleep and physical activity}

\begin{figure}[ht!]
    \centering
    \includegraphics[width=\columnwidth]{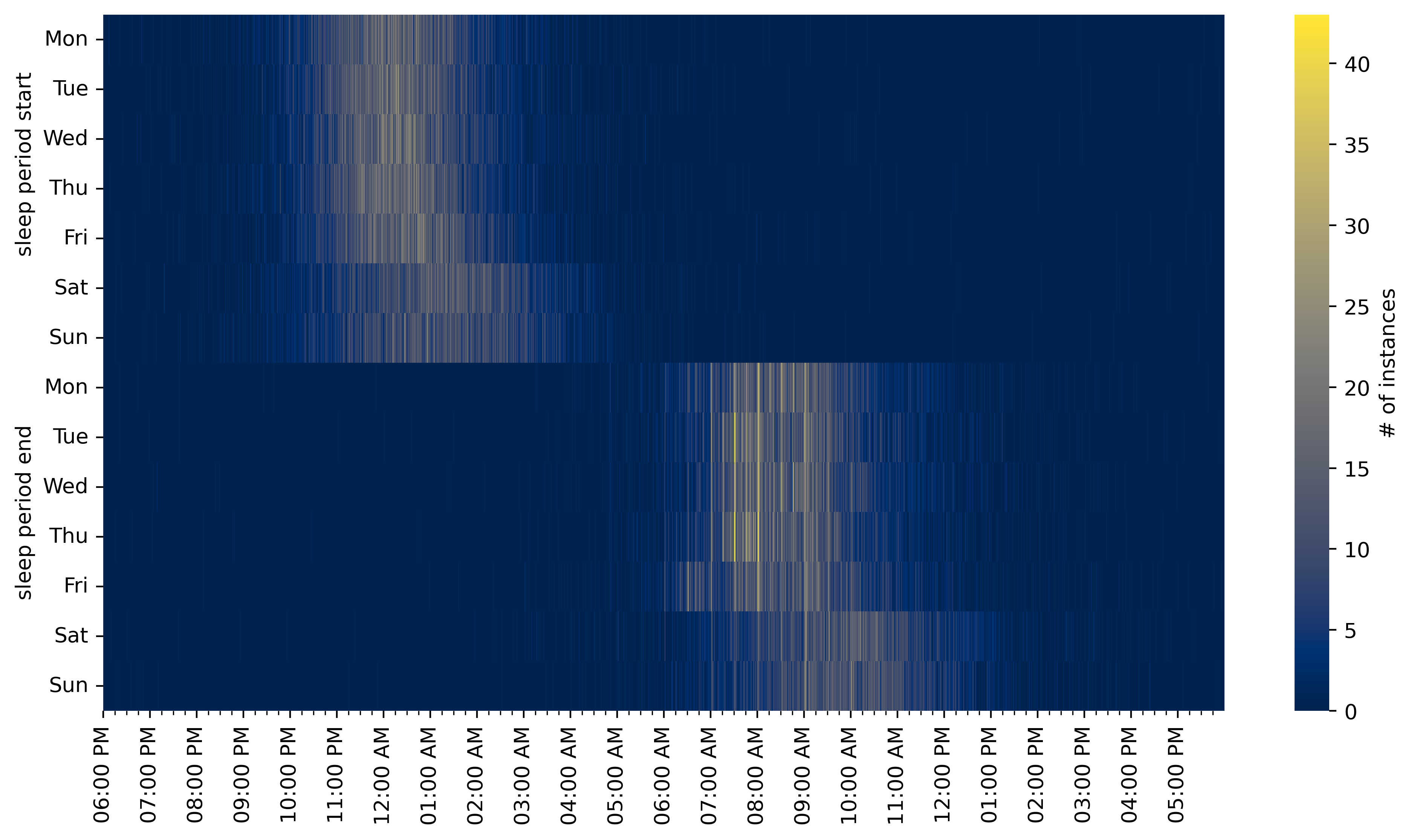}
    \caption{\textbf{Bedtimes and get-up times.} The heatmap shows the number of recorded start and end times of sleep periods at a 1-minute resolution. Note the presence of bright vertical lines for get-up times that are absent in bedtimes, suggesting the use of alarm clocks. The delay in bedtimes and get-up times during the weekends is also evident. Thanksgiving break is excluded in the plot.}
    \label{fig:sleep-heatmap}
\end{figure}

Figure~\ref{fig:sleep-heatmap} shows the bedtimes and get-up times for all users and all days.
Note the distinct bright vertical lines in get-up time that are not present in bedtime, indicating that more users share specific get-up times than bedtimes.
The most popular get-up times are 1 minute away from the quarter marks on a clock, i.e., :01, :16, :31, :46, suggesting the use of alarm clocks.
In contrast, shared bedtimes are less common.
Bedtimes and get-up times both occur later during the weekends than the weekdays.

\begin{figure}[ht!]
    \centering
    \includegraphics[width=\columnwidth]{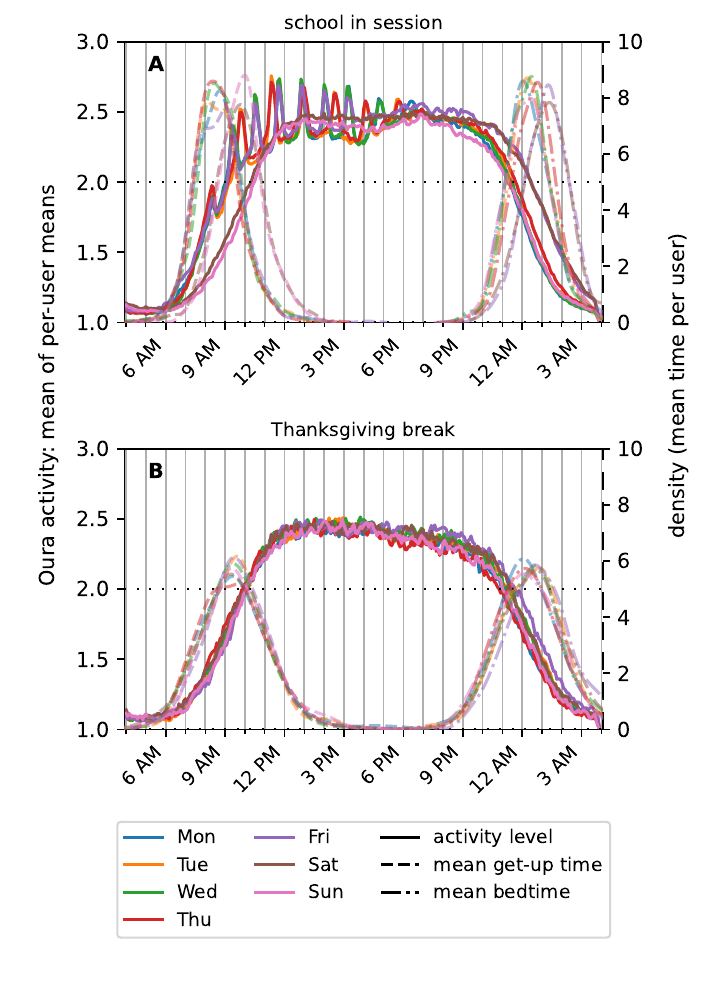}
    \caption{\textbf{Weekly patterns in sleep and activity.} The solid lines give the mean of the per-user means of the Oura activity level time series, disregarding non-wear periods. The ordinal activity level categories based on METs are reported as integers: 1=rest, 2=inactive, 3=low intensity, 4=medium intensity, 5=high intensity. The dashed and dash-dotted lines indicate the distribution of the mean bedtimes and get-up times, respectively, per user per day of the week.}
    \label{fig:weekly-sleep-activity}
\end{figure}

Figure~\ref{fig:weekly-sleep-activity} shows the distributions of the bedtimes and get-up times and the mean activity level across all users for each day of the week.
When school is in session (Figure~\ref{fig:weekly-sleep-activity}a), the activity time series during weekdays are characterized by spikes.
Further, these spikes coincide for Mondays, Wednesdays, and Fridays, as well as for Tuesdays and Thursdays.
This is consistent with class scheduling in the university where the participants are based, with classes scheduled either Mon-Wed-Fri for 50 minutes or Tues-Thurs for 75 minutes.
In contrast, these spikes are absent during the weekends as well as during Thanksgiving break (Figure~\ref{fig:weekly-sleep-activity}b).
Both bedtimes and get-up times are later during the weekends compared to weekdays, suggesting widespread social jetlag among the participants which is consistent with previous findings on college students~\cite{lund_sleep_2010}.
Consequently, activity levels both increase and wind down later in the day during the weekends.

\begin{figure}[ht!]
    \centering
    \includegraphics[width=\columnwidth]{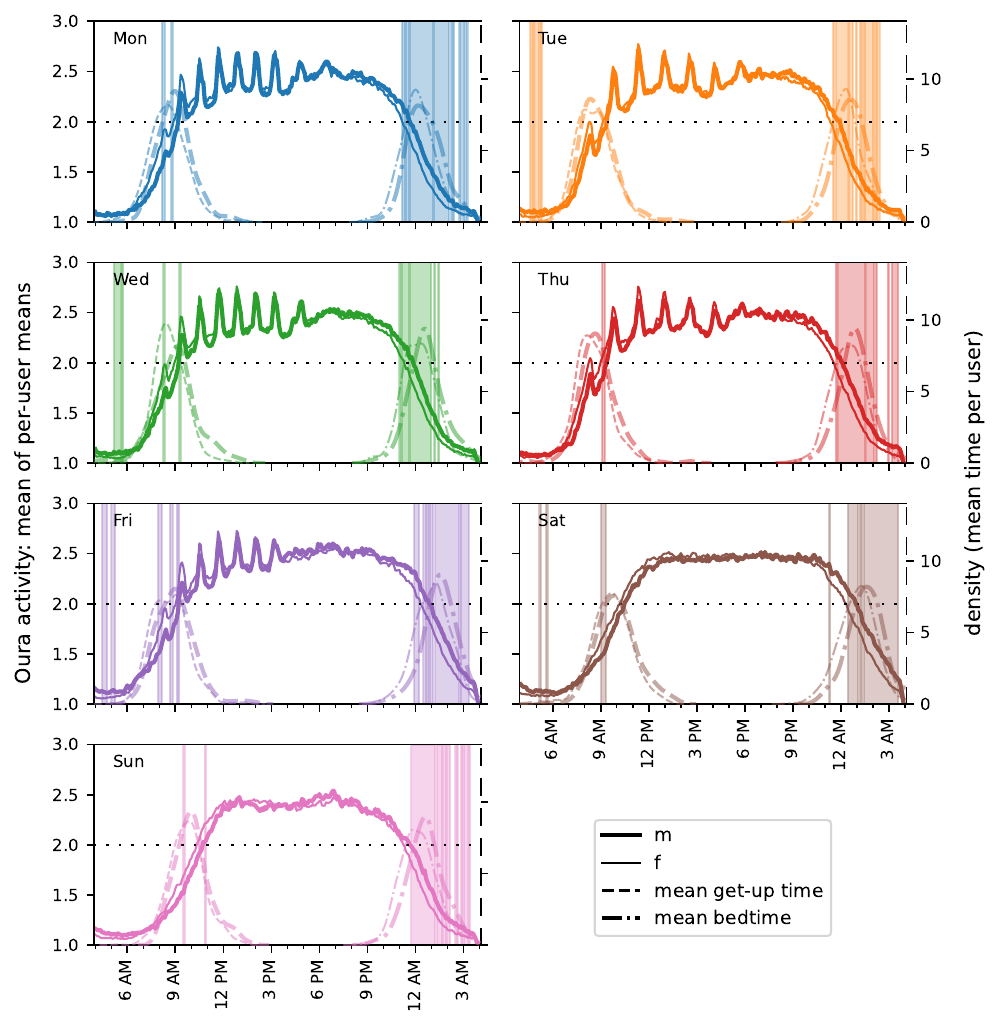}
    \caption{\textbf{Comparison of activity time series of males and females.} Periods in the activity time series with differences statistically significant at $\alpha=0.05$ according to statistical parametric mapping (SPM) are shaded. For males and females, these differences are concentrated around bedtime. Regardless, the daytime weekday spikes in the activity time series are at the same times for both males and females.}
    \label{fig:actts_m_vs_f}
\end{figure}

While these general trends persist when we subset the population by gender or mental health indicators, we observe differences in specific details.
Males go to bed and get up later in the day but have a lower sleep efficiency (Table S2).
This is supported by differences in the activity time series across males and females: SPM reveals differences in activity levels are most prominent during bedtime.
Despite this difference in bedtime habits, the spikes in activity during the daytime on weekdays coincide for males and females (Figure~\ref{fig:actts_m_vs_f}).
Impairment, on the other hand, is not associated with differences in activity levels, even if we consider it only among a subset of a specific gender 
(Figures S4--S6).

\subsection{Social jetlag and activity}

\begin{table*}[ht]
\caption{\textbf{Linear regression model predicting the mean midpoint of sleep during Thanksgiving week using gender and impairment status.} Gender and impairment are used as predictors for the MSF in a linear regression model.}
\label{tab:chronotype_reg}
\begin{tabular}{|l|l|r|r|r|r|}
\hline
\textit{Response}          & \textit{Fixed Effect}     & \textit{Coeff} & \textit{2.5 CI} & \textit{97.5 CI} & \textit{p-value} \\ \hline
Sleep midpoint, MSF (hrs from 12MN)   & Gender - male  & 0.472  & 0.244  & 0.700 & \textbf{\textless{}0.001} \\
 & Gender - non-binary  &  0.209  & -0.215  & 0.633 & 0.333 \\
 & Impairment  &  0.273  & 0.056  & 0.489 & \textbf{0.014} \\
 \hline
 \end{tabular}
\end{table*}

\begin{table*}[ht]
\caption{\textbf{Sleep period midpoint and bedtime during free days.} Each row corresponds to a linear regression model relating the mean sleep period midpoint and the mean of a single sleep-related quantity, both taken during Thanksgiving week. The sleep period midpoint (or the MSF) is related to the mean start and end of the sleep period but not the sleep period duration. This is graphically demonstrated in Figure~\ref{fig:chronotype_dist}B.}
\label{tab:chronotype_reg_duration}
\begin{tabular}{|l|l|r|r|r|r|}
\hline
\textit{Response}          & \textit{Fixed Effect}     & \textit{Coeff} & \textit{2.5 CI} & \textit{97.5 CI} & \textit{p-value} \\ \hline
Sleep midpoint, MSF (hrs from 12MN) & Sleep period start  & 0.776  & 0.746  & 0.806 & \textbf{\textless{}0.001} \\ \hline
Sleep midpoint, MSF (hrs from 12MN) 
 & Sleep period end  &  0.841  & 0.804  & 0.878 & \textbf{\textless{}0.001} \\ \hline
 Sleep midpoint, MSF (hrs from 12MN) 
 & Sleep period duration  &  -0.09  & -0.192  & 0.012 & 0.084 \\
 \hline
 \end{tabular}
\end{table*}

\begin{figure}[ht]
    \centering
    \includegraphics[width=\columnwidth]
    {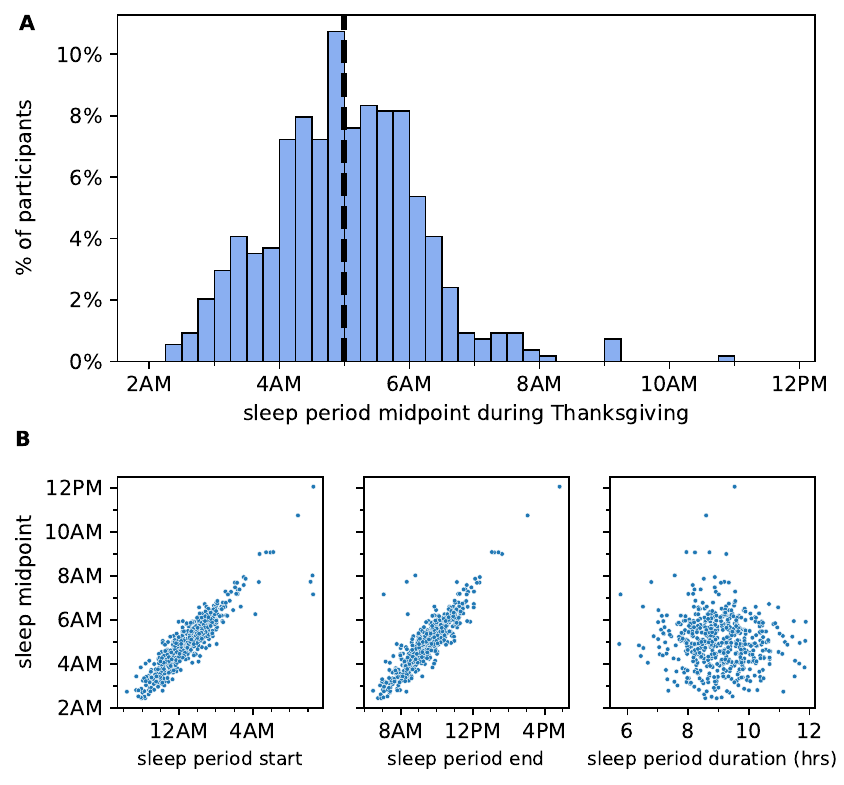}
    \caption{\textbf{Chronotype and bedtime.} (a) The chronotype is quantitatively represented by the mean time of the midpoint of the sleep period during Thanksgiving week. The median, around 5AM, is shown with a dashed line. (b) The sleep period midpoint is affected more by the start and end of the sleep period but not by the duration.}
    \label{fig:chronotype_dist}
\end{figure}

\begin{figure}[ht!]
    \centering
    \includegraphics[width=\columnwidth]
    {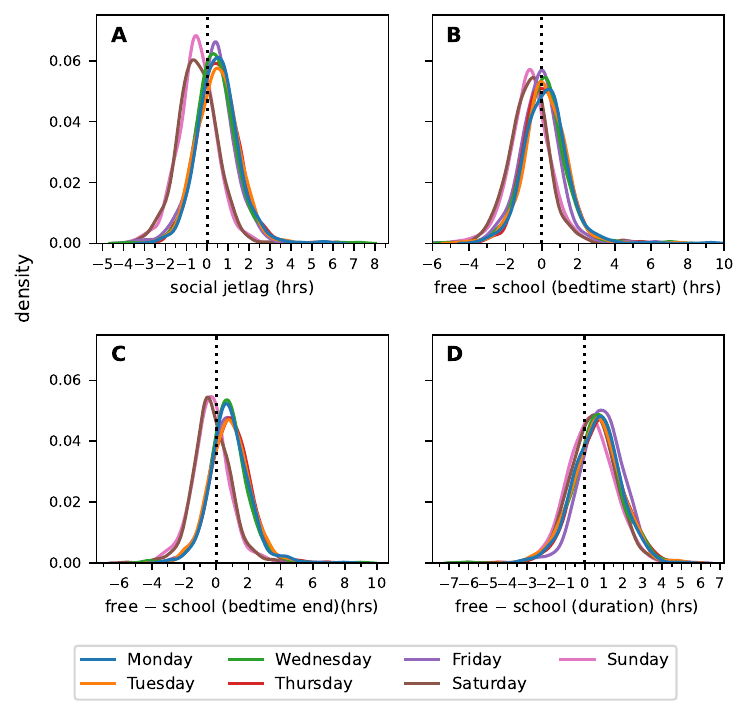}
    \caption{\textbf{Social jetlag for weekdays and weekends.} These are density plots of the differences in the per-user means of bedtime-related quantities for every day of the week between free days (Thanksgiving week) and school days (excluding Thanksgiving week). The days of the week correspond to the day of the end of the sleep period, and the dashed line indicates no difference between free days and school days. Generally speaking, (a) social jetlag, computed as the difference between the midpoints of sleep periods of free and work days ($SJL=MSF-MSW$), is positive in the weekdays but negative in the weekends. When school is in session, (b) students sleep a little earlier than on Thanksgiving week from Sundays to Wednesday, a little later on Thursdays, but much later on Fridays and Saturdays (note that the day of the week in the plot corresponds to the day upon waking up). (c) They get up earlier on weekdays than on Thanksgiving week, but later on the weekends. (d) However, for both weekdays and weekends when school is in session, students are getting less sleep than they do on Thanksgiving week.}
    \label{fig:jetlag_kdeplots}
\end{figure}

\begin{table*}[]
\caption{\textbf{Differences in bedtime-related quantities between Thanksgiving week and school weeks.}}
\label{tab:sleep_tgv_vs_school}
\begin{tabular}{|r|rrr|}
\hline
\multicolumn{1}{|l|}{}      & \multicolumn{3}{c|}{\textit{mean in Thanksgiving $-$ mean in other weeks (minutes)}}                                    \\ \hline
\textbf{}                   & \multicolumn{1}{r|}{\textbf{Bedtime start}} & \multicolumn{1}{r|}{\textbf{Bedtime end}} & \textbf{Duration} \\
\textit{day bedtime ends}   & \multicolumn{1}{r|}{\textbf{}}              & \multicolumn{1}{r|}{\textbf{}}            & \textbf{}         \\ \hline
\textit{\textbf{Monday}}    & \multicolumn{1}{r|}{12.27}                  & \multicolumn{1}{r|}{50.16}                & 37.92             \\ \hline
\textit{\textbf{Tuesday}}   & \multicolumn{1}{r|}{10.42}                  & \multicolumn{1}{r|}{49.44}                & 37.50             \\ \hline
\textit{\textbf{Wednesday}} & \multicolumn{1}{r|}{5.81}                   & \multicolumn{1}{r|}{44.65}                & 37.02             \\ \hline
\textit{\textbf{Thursday}}  & \multicolumn{1}{r|}{9.15}                   & \multicolumn{1}{r|}{53.59}                & 42.55             \\ \hline
\textit{\textbf{Friday}}    & \multicolumn{1}{r|}{-8.52}                  & \multicolumn{1}{r|}{46.28}                & 54.68             \\ \hline
\textit{\textbf{Saturday}}  & \multicolumn{1}{r|}{-41.60}                 & \multicolumn{1}{r|}{-15.19}               & 25.08             \\ \hline
\textit{\textbf{Sunday}}    & \multicolumn{1}{r|}{-39.00}                 & \multicolumn{1}{r|}{-17.98}               & 19.57             \\ \hline
\end{tabular}
\end{table*}

Our sample is composed mainly of late-night chronotypes with a median MSF (from Thanksgiving week) of 5AM (Figure~\ref{fig:chronotype_dist}A).
Being male ($p<0.001$) and reporting impairment due to anxiety or depression ($p=0.014$) are both associated with later chronotypes (Table~\ref{tab:chronotype_reg}).
Late-night chronotypes are characterized by both going to bed and getting up from bed later in the day (Figure~\ref{fig:chronotype_dist}B).
The relationship between MSF and the mean sleep duration during Thanksgiving week is not statistically significant (Table~\ref{tab:chronotype_reg_duration}), consistent with previous studies~\cite{roenneberg_epidemiology_2007}.
The mean difference of the sleep period midpoint from the MSF outside of Thanksgiving week is positive (i.e., earlier midpoint of sleep) during weekdays and negative (i.e., later midpoint of sleep) during weekends (Figure~\ref{fig:jetlag_kdeplots}A).
Compared to Thanksgiving week, bedtime starts earlier when the following day is from Monday to Thursday.
It starts a little later when the next day is a Friday, but by more than half an hour when the following day is a weekend (Figure~\ref{fig:jetlag_kdeplots}B and Table~\ref{tab:sleep_tgv_vs_school}).
As expected, bedtime ends earlier on weekdays than on weekends (Figure~\ref{fig:jetlag_kdeplots}C and Table~\ref{tab:sleep_tgv_vs_school}).
However, regardless of whether it is a weekday or a weekend, students sleep longer on Thanksgiving week than they do during school weeks (Figure~\ref{fig:jetlag_kdeplots}D and Table~\ref{tab:sleep_tgv_vs_school}).

Social jetlag is also higher for late-night chronotypes (Figure S7),
consistent with what is found in the literature~\cite{roenneberg_chronotype_2019}.
Gender and impairment due to anxiety or depression are not statistically significant predictors of social jetlag.
This result holds whether the day of the week (as weekday vs. weekend or as a categorical variable with 7 levels) is considered a fixed or a random effect (Table~\ref{tab:jetlag_demog_reg}).

\begin{table*}[ht!]
\caption{\textbf{Mixed-effects linear models predicting social jetlag from gender, impairment, and the day of the week.} This table shows the results of mixed-effects linear models for predicting social jetlag from gender, impairment, and the day of the week the sleep period ends. The random effects were the participant ID and the week number. Thanksgiving week was excluded in this analysis.}
\label{tab:jetlag_demog_reg}
\begin{tabular}{|l|l|r|r|r|r|}
\hline
\textit{Response}          & \textit{Fixed Effect}     & \textit{Coeff} & \textit{2.5 CI} & \textit{97.5 CI} & \textit{p-value} \\ \hline
Social jetlag (hrs)  & Gender - male  & -0.026  & -0.189  & 0.136 & 0.751 \\
 & Gender - non-binary  &  -0.170  & -0.472  & 0.132 & 0.271 \\
 & Impairment  &  -0.039  & -0.193  & 0.115 &  0.619 \\
& Day of week (weekday=1, weekend=0)  &  0.924  & 0.892  & 0.957 &  \textbf{\textless{}0.001} \\
 \hline
 Social jetlag (hrs)  & Gender - male  & -0.027  & -0.190  & 0.136 & 0.745 \\
 & Gender - non-binary  &  -0.170  & -0.472  & 0.132 & 0.271 \\
 & Impairment  &  -0.039  & -0.194  & 0.115 &  0.618 \\
& Day of week - Tue  &  -0.019  & -0.074  & 0.036 &  0.503 \\
& Day of week - Wed  &  -0.115  & -0.170  & -0.060 &  \textbf{\textless{}0.001} \\
& Day of week - Thu  &  -0.005  & -0.060  & 0.050 &  0.856 \\
& Day of week - Fri  &  -0.197  & -0.252  & -0.141 &  \textbf{\textless{}0.001} \\
& Day of week - Sat  &  -1.000  & -1.056  & -0.945 &  \textbf{\textless{}0.001} \\
& Day of week - Sun  &  -0.983 & -1.039  & -0.928 &  \textbf{\textless{}0.001} \\
 \hline
 
\end{tabular}
\end{table*}

Social jetlag also affects activity, with positive social jetlag being associated with more active calories the following day even after controlling for the day of the week (weekday vs. weekend).
Being male is also associated with higher active calories, while weekends, longer bedtimes the night prior and impairment due to anxiety or depression are associated with reduced active calories (Table~\ref{tab:calactive_reg}).

\begin{table*}[ht!]
\caption{\textbf{Mixed-effects linear model predicting active calories expended.} Gender, impairment, social jetlag and bedtime duration for the night before, and the day of the week the sleep period ends (whether weekday or weekend) were considered as fixed effects. The random effects were the participant ID and the week number. Thanksgiving week was excluded in this analysis.}
\label{tab:calactive_reg}
\begin{tabular}{|l|l|r|r|r|r|}
\hline
\textit{Response}          & \textit{Fixed Effect}     & \textit{Coeff} & \textit{2.5 CI} & \textit{97.5 CI} & \textit{p-value} \\ \hline
Active calories (kcal, log-transformed)  & Gender - male  &  0.106  & 0.078  & 0.135 & \textbf{\textless{}0.001} \\
 & Gender - non-binary  &  0.024  & -0.029  & 0.077 & 0.371 \\
 & Impairment  &  -0.027  & -0.054  & -0.000 &  \textbf{0.048} \\
 & Social jetlag (hrs)  &  0.020  & 0.017  & 0.022 &   \textbf{\textless{}0.001} \\
  & Bedtime duration (hrs)  &  -0.019  & -0.021  & -0.017  &   \textbf{\textless{}0.001} \\
& Day of week (weekday=1, weekend=0)  & 0.062  & 0.056  & 0.069 &  \textbf{\textless{}0.001} \\
 \hline
 
\end{tabular}
\end{table*}
\section{Summary and Discussion}
\label{sec:discussion}

To understand collective sleep and activity habits of young adults who have recently started college, we monitored more than 500 first-year students in the same university through surveys and wearable devices for a continuous period of 8 weeks in a single semester.
In contrast to prior studies that only use the daily sleep and/or activity measures, we utilize fine temporal resolution data from the wearable device to observe shorter events within the course of a day.

We observed a high compliance rate for both daytime and nighttime Oura ring wear, with at least 70\% of our sample having sleep and activity data for a given day.
This is consistent with those observed for college students using different wearable devices~\cite{purta_experiences_2016}.
Participants wore the ring consistently throughout the day, with dips in wear concentrated once in the morning and once in the evening.
The non-wear period in the morning shifts later in the day during the weekends, while the non-wear period in the evening remains at the same time for both weekdays and weekends.
As both the start and end of the sleep period are delayed during the weekends, the morning non-wear period appears to be related to the end of the sleep period, while the evening non-wear period may be related to an evening activity that does not relate to sleep.
The high rate of compliance shows that the Oura ring can be used to monitor the effects of interventions on the various physiological measures taken by the Oura ring.
However, the timing of non-wear suggests that some activity shortly after waking up and in the evening may be missed in data collection.

From the sleep data, we found 
commonality in the get-up times which is likely due to alarm clock use.
In contrast, there is more variety in bedtimes, indicating less structure in going to bed than getting up from bed.
Generally, participants had a 
natural late-night chronotypeonsistent with prior studies~\cite{lund_sleep_2010}, with the median sleep period midpoint time at around 5AM.
Being male and having mental health impairment were associated with later chronotypes.
Interestingly, social jetlag, or a deviation from a natural sleep pattern, is found not just for weekdays but also for weekends when school is in session, and is not associated with gender or mental health impairment.
School weekdays are characterized by earlier and shorter sleep periods than during school breaks, consistent with academic demands.
School weekends, on the other hand, have sleep periods that are later but still shorter than on school breaks.

While the weekday-weekend differences in sleep habits is well-known~\cite{mcmahon_persistence_2018}, that students sleep for longer during school breaks than school weekends implies that the sleep debt is not sufficiently recovered while school is in session.
Further study is needed to see if this chronic sleep debt is partly responsible for mental health struggles found among college students and whether changing school schedules would be beneficial to student health and performance~\cite{onyper_class_2012, vollmer_morningness_2013, yim_how_2023, yeo_early_2023}.

The collective activity time series showed synchronized spikes in activity coinciding with class schedules, likely corresponding to walking in between classes.
These spikes are observed at the same times regardless of gender or mental health impairment status.
With weekdays being associated with a higher active calorie expenditure when school is in session, walking between classes appears to significantly contribute to the daily physical activity of college students.
Prior studies have found that time spent in class is associated with decreased physical activity~\cite{wu_association_2023} but did not consider activity in between classes.
This is relevant in developing interventions to improve activity-related metrics particularly for students in smaller campuses.

The observation of the activity spikes during school weekdays is of particular interest for a number of reasons.
First, while the Oura ring is known for its high accuracy in recording sleep metrics~\cite{de_zambotti_sleep_2019, altini_promise_2021}, its performance for measuring activity is less well established~\cite{henriksen_polar_2022, niela-vilen_comparison_2022, kristiansson_validation_2023}.
With our results showing clear relationships between class schedules and activity, we have demonstrated that the Oura Gen3 ring activity data can be used to distinguish sedentary behavior (e.g., sitting while in class) from low- to moderate-intensity activity (e.g., walking between classes) at the collective level.

Second, our results show how analyzing wearable device data at finer temporal resolutions reveals patterns absent in daily aggregate measurements.
While differences between weekdays and weekends are also seen from the active calorie count, we are only able to see the spikes in activity because the Oura ring has a temporal resolution finer than the duration between classes.

Third, while our study focuses on collective behavior, the identification of synchronized activity solely from wearable device data also has implications for individual activity detection.
Each student had a maximum of seven (7) time series for each day of the week, each corresponding to a week in the study period (excluding Thanksgiving week).
Since Mon-Wed-Fri, Tues-Thurs, and Sat-Sun correspond to different routines, averaging individual time series where routine activity is expected results in more noise, making activity detection more challenging.
However, because synchronized activity is observable by taking the mean of several hundred time series belonging to different individuals, we may similarly detect regular time-bound activities from wearable device data of a \textit{single} individual if the data collection period is long enough.

Positive social jetlag for a sleep period, which corresponds to earlier sleep periods, was associated with increased active calories.
This is consistent with prior studies finding less intensity in physical activity with delayed bedtimes~\cite{shechter_delayed_2014, sempere-rubio_association_2022}.
Extended sleep duration is associated with lower active calories, which supports previous findings~\cite{pesonen_sleep_2022}.
We infer that sleeping for longer reduces the amount of time one can be active, and the participants do not increase their active calorie expenditure enough to compensate for this.
Weekends were also associated with lower active calories.
Taken altogether, these indicate that active calorie expenditure among college students may be less affected by deviations from natural circadian rhythms and more by the timing associated with activities.
Higher physical activity may be due to routines, such as walking in between classes, that are only done in the daytime, while evening and weekend habits may be associated with more sedentary activities.

We have shown how the high temporal resolution data from wearable devices can be used to capture collective sleep and activity patterns. 
The relatively homogeneous demographic profile of our sample and the shared academic context simplifies the interpretation of our findings and facilitates developing appropriate interventions.
For a larger population, this may not be as straightforward.
For example, while school schedules allow us to better estimate natural sleep habits through school breaks, this will be more challenging to measure for other stages of life.
However, it would be equally interesting to see if collective behaviors could still emerge and whether fine temporal resolution data can reveal patterns that would not be present in coarse-grained data.
Similarly, we can also look at how sleep and activity patterns vary for groups with different shared contexts, such as athletes or retirees.

Another avenue for future work is to fully utilize the gamut of information from wearable devices.
In this work, we focused mainly on activity levels and the timing of sleep periods.
However, the Oura ring also provides other information, such as heart rate, heart rate variability, respiratory rate and changes in skin temperature, among others.
Building from our prior work~\cite{bloomfield_predicting_2024, fudolig_two_2024}, we hope to understand these physiological measures in relation to routines.

\acknowledgments

This study was funded by a grant from MassMutual. The manuscript benefited from helpful discussions with Thomas Sorensen and Alec Beauregard.

\clearpage

\end{document}